# Realization of a coupled-mode heat engine with cavity-mediated nanoresonators

Jiteng Sheng[1,2], Cheng Yang[1], and Haibin Wu[1,2,3,*]

[1]State Key Laboratory of Precision Spectroscopy, East China Normal University, Shanghai 200062, China

[2]Collaborative Innovation Center of Extreme Optics, Shanxi University, Taiyuan 030006, China

[3]Shanghai Research Center for Quantum Sciences, Shanghai 201315, China

*hbwu@phy.ecnu.edu.cn

**Abstract:** We report an experimental demonstration of a coupled-mode heat engine in a two-membrane-in-the-middle cavity optomechanical system. The normal mode of the cavity-mediated strongly coupled nanoresonators is utilized as the working medium and an Otto cycle is realized by extracting work between two phononic thermal reservoirs. The heat engine performance is characterized in both normal mode and bare mode pictures, which reveals that the correlation of two membranes plays a significant role during the thermodynamic cycle. Moreover, a straight-twin nanomechanical engine is implemented by engineering the normal modes and operating two cylinders out-of-phase. Our results demonstrate an essential class of heat engine for the first time in cavity optomechanical systems and provide a novel platform for investigating heat engines of interacting subsystems in small scales with controllability and scalability.

# INTRODUCTION

Heat engine, as an essential achievement in thermodynamics, has regained significant attention in the non-equilibrium regime with the developments in nanotechnology and laser cooling. Heat engines at nano/microscales or single-atom levels have been experimentally realized with a single trapped ion [1,2], nano/micro-resonators [3-7], nitrogen vacancy centers [8], cold atoms [9,10], and nuclear spins [11,12]. On the other hand, optomechanics, due to the flexible controllability and extremely low decoherence of mechanical oscillators, has witnessed tremendous achievements in exploring quantum physics in mesoscopic or macroscopic scales [13-18] and ultrasensitive metrology [19-24]. Recently, significant progress in fabrication of mechanical devices at micro- and nanoscales has made the optomechanical system as an excellent candidate for studying non-equilibrium thermodynamics [25-28]. A hallmark example is to study the stochastic [4-7] and quantum [8-12] heat engines in such a system. Although the optomechanical system has several appealing advantages, e.g. it is a truly mechanical system and has the ability to operate deep in the quantum regime, and several theoretical models for heat engines in optomechanics have been proposed [29-38], the experimental studies of heat engine in cavity optomechanical systems remain elusive so far.

In this work, we demonstrate a coupled-mode stochastic heat engine in a multimode optomechanical system with two nanomechanical membranes inside an optical cavity [39-41]. Although the strongly coupled oscillators have been intensively investigated as an important model of stochastic or quantum heat engines in many theoretical works [42-45], the experimental demonstration has not been reported. Here, we extend the proposal [29] and realize a heat engine based on cavity-mediated strongly interacting nanomechanical membranes for the first time. The normal mode of two nanomechanical resonators is utilized as the working medium, and an Otto cycle is implemented by controlling the frequency of membrane and the phononic thermal bath. We have developed a method to analyze the work and efficiency of such a coupled-mode engine in the normal mode and bare mode pictures, incorporating with the technique of single trajectory real-time measurement. Remarkably, the correlation of two membranes plays an important role and performs considerable work in the thermodynamic cycle. Moreover, such a multimode system can be straightforwardly extended to multi-cylinder heat engines. A straight-twin nanomechanical heat engine is realized by engineering the normal modes and exploiting two normal mode branches alternatively in the same thermodynamic cycle. The realization of such a coupled-mode heat engine with optomechanics extends the nanomechanical heat engines to multipartite systems with high flexibility, and provides an opportunity to study more interesting phenomena in non-equilibrium thermodynamics with interacting systems.

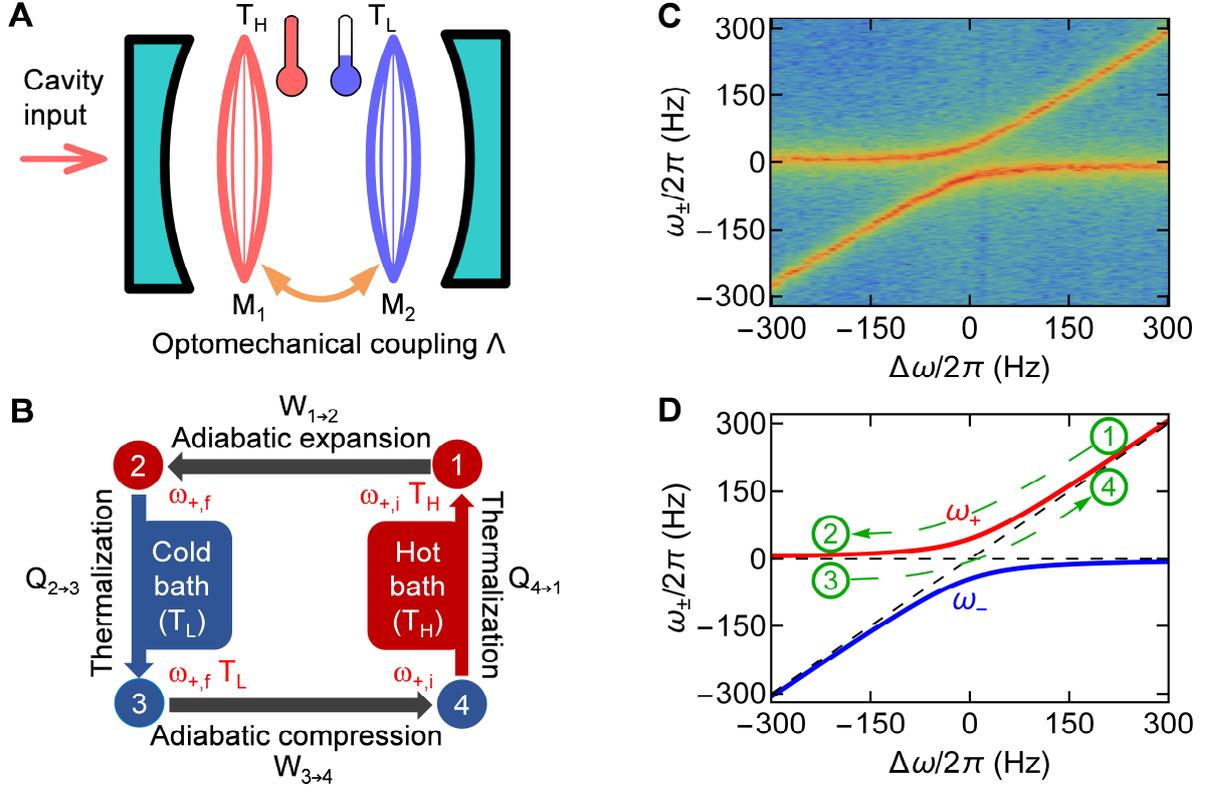

Fig. 1. A coupled-mode heat engine with strongly interacting nanomechanical membranes. (A) Two optomechanically coupled nanomechanical membranes ($M_1$ and $M_2$) with one interacting with a room temperature thermal bath ($T_L$) and the other with a high temperature one ($T_H$). (B) Schematic of the Otto cycle associated with the upper normal mode. (C) Measured thermomechanical noise spectra as a function of $\Delta\omega$, which indicates the anti-crossing behavior in the strong coupling regime. The effective coupling strength is $\Lambda=2\pi\times40$ Hz, which is larger than the damping rates of membranes ($\gamma_1=2\pi\times6$ Hz and $\gamma_2=2\pi\times12$ Hz). (D) Corresponding theoretical plot of the normal mode frequencies. The black dashed lines represent the bare modes of membranes ($M_1$ and $M_2$), and the red and blue solid curves are the normal modes of the effective coupled system ($M_+$ and $M_-$). Four strokes of an Otto cycle operated on the upper normal mode are denoted, in correspondence with the schematic shown in Fig. 1B. The strokes from status 1 to 2 and from 3 to 4 correspond to the "adiabatic" processes. The strokes from status 2 to 3 and from 4 to 1 correspond to the isochoric processes.

## RESULTS AND DISCUSSION

### Experimental setup

The experimental setup of the coupled-mode nanomechanical heat engine is similar to the one used in Refs. [28,46], as shown in Fig. 1A. Two spatially-separated silicon nitride nanomechanical membranes are placed inside a Fabry-Perot cavity. One membrane is in contact with a room temperature thermal bath and the other is driven by a white noise to obtain the high temperature thermal bath [28]. The motions of membranes are monitored by two weak probe laser fields (not shown in the figure). The vibrational (1,1) modes are utilized in the experiment, which are nearly degenerate with eigenfrequencies $\omega_m \approx 2\pi \times 400$ kHz (m=1,2). Piezos are used to precisely control the frequencies of membranes [41]. The cavity is driven by a red-detuned laser field, which interacts with both membranes simultaneously due to the dynamical backaction, and consequently two individual eigenmodes (bare modes) of membranes are effectively coupled by the cavity field.

### Hamiltonian description

The interaction Hamiltonian of such a composite system of two membranes interacting with a common cavity field is given by $\hat{H}_{int} = -\hbar \sum_{i=1,2} g_i \hat{a}^\dagger \hat{a} (\hat{b}_i^\dagger + \hat{b}_i)$, where $\hat{a}$ and $\hat{b}_i$ are the annihilation operators for cavity and mechanical modes, respectively. $g_i$ is the optomechanical coupling rate of each membrane i. After adiabatically eliminating the cavity mode, the system can be effectively described by a phonon-phonon coupling Hamiltonian [28,47], $\hat{H}_{eff} = \hbar \sum_{i=1,2} \Omega_i \hat{b}_i^\dagger \hat{b}_i + \hbar \Lambda (\hat{b}_2^\dagger \hat{b}_1 + \hat{b}_1^\dagger \hat{b}_2)$, where $\Omega_i = \omega_i - i\gamma_i/2 - \Lambda$ and $\Lambda = g_1 g_2 \chi_{eff}$ is the effective coupling strength between two membranes. Here, the membranes are assumed to have the same optomechanical coupling strength ($|g_1|=|g_2|$), and $\chi_{eff}$ is the effective susceptibility introduced by the intracavity field [48,49]. The laser frequency is detuned far-off the cavity resonance, and the interacting of membranes is dominated by a conservative coupling. In the strong coupling regime ($\Lambda \gg \gamma_{1,2}$), the normal mode splitting appears. The anti-crossing behavior of the noise spectrum is clearly observed by varying the frequency detuning of membranes ($\Delta\omega = \omega_1 - \omega_2$), as presented in Fig. 1C. Figure 1D is the corresponding theoretical interpretation, in which the black dashed lines represent the bare modes of membranes, and the red and blue solid curves are the normal modes of the effective coupled system, i.e., $\omega_\pm = \Delta\omega/2 \pm \sqrt{\Delta\omega^2/4 + \Lambda^2}$, by ignoring the imaginary part and a constant energy shift. It is worth noting that the strong coupling is realized between two mechanical modes mediated by the optical field, rather than the case of strong optomechanical coupling regime, where the optomechanical coupling rate is larger than the cavity bandwidth [48].

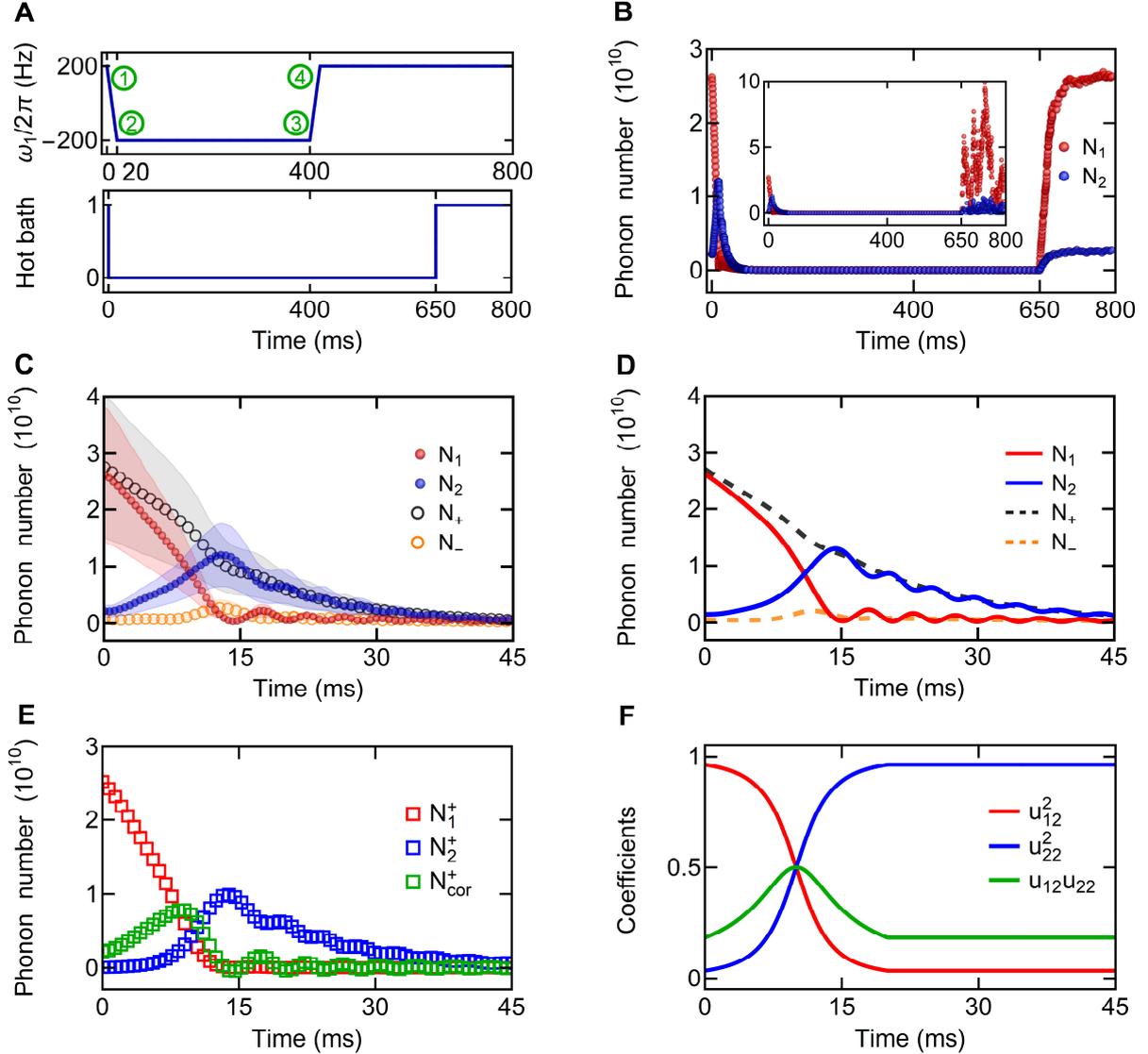

**Fig. 2. Dynamics of phonon numbers and correlation in a thermodynamic cycle.** (A) The frequency of $M_1$ and the high temperature thermal bath as a function of time during a thermodynamic cycle. The status numbers are marked according to Fig. 1D. 1 = on and 0 = off for the hot bath. (B) The measured average phonon numbers of bare modes over 250 trajectories. The inset shows a typical single trajectory of the cycle. (C) The measured phonon numbers during the expansion stroke (part of the isochoric stroke is included). The red and blue dots presents the phonon numbers of bare modes, i.e. $N_1$ and $N_2$. The grey and orange circles are the phonon numbers of normal modes, i.e. $N_+$ and $N_-$. The error bars (shaded

regions) are the standard deviations. The fluctuations of phonon numbers obey the thermal Boltzmann distribution. (D) The corresponding theoretical simulation. (E) Different components of N$_+$ during the expansion stroke. $N_1^+$, $N_2^+$, and $N_{cor}^+$ represent the terms of the phonon numbers and correlation for the upper normal mode, which are $u_{12}^2 \hat{b}_1^\dagger \hat{b}_1$, $u_{22}^2 \hat{b}_2^\dagger \hat{b}_2$, and $u_{12} u_{22} \left( \hat{b}_1^\dagger \hat{b}_2 + \hat{b}_2^\dagger \hat{b}_1 \right)$, respectively. (F) The corresponding coefficient of each component for N$_+$.

**Experimental results**

The basic principle of such a coupled-mode heat engine works similarly to the previous theoretical proposal [29]. We first consider the heat engine operates along a single normal mode and focus on the upper normal mode. Here the upper normal mode branch is used as the working medium, as shown in Fig. 1D. The frequency of the normal mode plays the role of the volume of the working medium. The expansion can be realized by reducing the normal mode frequency ω$_+$, which is controlled by Δω. The heat engine is based on an Otto cycle with two "adiabatic" and two isochoric processes. The schematic of the Otto cycle associated with the upper normal mode is shown in Fig. 1B. The energy changes of the four strokes for the upper normal mode are denoted as W$_{1\to2}$, Q$_{2\to3}$, W$_{3\to4}$, and Q$_{4\to1}$, respectively. The total work done by the upper normal mode for an ideal Otto cycle is W=W$_{1\to2}$+W$_{3\to4}$=ℏ(ω$_{+,i}$−ω$_{+,f}$)(⟨N$_+$⟩$_f$−⟨N$_+$⟩$_i$), where ⟨N$_+$⟩$_i$=k$_B$T$_H$/ℏω$_{+,i}$ and ⟨N$_+$⟩$_f$=k$_B$T$_L$/ℏω$_{+,f}$ are the mean phonon numbers of upper normal mode at status 1 and 3, respectively. The heat that the upper normal mode consumes from the hot bath is Q$_{4\to1}$= ℏω$_{+,i}$(⟨N$_+$⟩$_i$−⟨N$_+$⟩$_f$). The work efficiency of the heat engine for the upper normal mode is defined as η=W/Q$_{4\to1}$.

The thermodynamic cycle starts at the status 1 (see Fig. 1D). At t=0, M$_1$ contacts with the high temperature thermal bath, while M$_2$ contacts with the room temperature thermal bath. The initial frequency detuning is Δω$_i$ ~ 2π×200 Hz, and the phonon number of upper normal mode is dominated by the phonon number of bare mode M$_1$, i.e., $\langle \hat{B}_+^\dagger \hat{B}_+ \rangle \approx \langle \hat{b}_1^\dagger \hat{b}_1 \rangle$. Subsequently, the high temperature thermal bath is turned off, the frequency of M$_1$ (ω$_1$) linearly sweeps from Δω$_i$ to the final frequency detuning Δω$_f$ ~ −2π×200 Hz within a time duration of 20 ms, and the frequency of M$_2$ (ω$_2$) keeps constant, i.e. the status 1 to 2, as illustrated in Fig. 1D. Consequently, the upper normal mode frequency changes from ω$_{+,i}$ ≈ 2π×200 Hz to ω$_{+,f}$ ≈ 0 Hz. The time sequences of ω$_1$ and the thermal bath are presented in Fig. 2A. This process corresponds to the expansion stroke of the classical heat engine, with a decrease of the upper normal mode frequency $\omega_+$. This is the key stroke of the cycle that the heat engine performs work to the environment during this process.

In our current setup, the work is performed by the nanomechanical membranes against the environment, i.e., the piezo and substrate, in the form of stress, which is dissipated by exciting unconfined mechanical modes over the system [7]. In practice, it is important to find a way to utilize the work via a flywheel or battery, for example, a flywheel can be directly attached to the end mirror in a optomechanical piston engine [29], the coupling between the engine and flywheel can be realized via a spin-dependent optical dipole force [2], and the work output can be stored in the quantum coherence and utilized by attaching an electrical load [37]. In order to harvest the work in our situation, the mechanical oscillators can be designed by integrating an additional oscillator in the system, so that the expansion of the membrane can drive a third mechanical oscillator coherently, which is similar to the method used in Ref. [3].

The stroke from status 2 to 3 is implemented in the constant frequencies and corresponds to the isochoric stroke. The upper normal mode experiences a full thermalization to a low phonon population at room temperature in this stroke. The process of status 3 to 4 is the compression process, with $\omega_1$ sweeping back to its initial value ($\Delta\omega_f \rightarrow \Delta\omega_i$). During this process, the environment does work to the heat engine, however, this is negligible due to the upper normal mode has been thermalized at the room temperature. In the process of status 4 to 1 (isochoric stroke), the high temperature thermal bath is switched on, and the upper normal mode is thermalized to a high phonon population state, which returns to the origin of the thermodynamic cycle. The measured phonon numbers of bare modes during the whole thermodynamic cycle are plotted in Fig. 2B, which are average over 250 cycles. A typical single trajectory of the cycle is shown in the inset of Fig. 2B. The stochastic property of the engine is dominated by the thermalization process of status 4 to 1. The fluctuation of the thermal population during the thermalization process leads to different initial phonon numbers for the expansion stroke in each cycle. Consequently, the work done in each cycle varies, which obeys an exponential function (see supplementary materials).

To better understand the expansion stroke, the detail of this process is illustrated in Fig. 2C. In the experiment, the mechanical displacements of bare modes, i.e. the amplitudes of the membrane vibrations, are measured in real-time. By diagonalizing the effective Hamiltonian, the normal modes operators can be written as $\hat{B}_+ = u_{12}\hat{b}_1 + u_{22}\hat{b}_2$ and $\hat{B}_- = u_{11}\hat{b}_1 + u_{21}\hat{b}_2$, where $\{\{u_{11}, u_{12}\}, \{u_{21}, u_{22}\}\}$ is the transform matrix. Thus, the phonon numbers of normal modes can be expressed as $N_+ = u_{12}^2 \hat{b}_1^\dagger \hat{b}_1 + u_{22}^2 \hat{b}_2^\dagger \hat{b}_2 + u_{12} u_{22} \left( \hat{b}_1^\dagger \hat{b}_2 + \hat{b}_2^\dagger \hat{b}_1 \right)$ and $N_- = u_{11}^2 \hat{b}_1^\dagger \hat{b}_1 + u_{21}^2 \hat{b}_2^\dagger \hat{b}_2 + u_{11} u_{21} \left( \hat{b}_1^\dagger \hat{b}_2 + \hat{b}_2^\dagger \hat{b}_1 \right)$, which are plotted in Fig. 2C. Figure 2D shows the corresponding theoretical simulation, which agrees very well with the experimental results. In Fig. 2E, the components for the phonon number of upper normal mode are illuminated. It is clearly seen that the population of the bare mode $M_1$ transfers to the mode $M_2$ and their correlation during the process, and the correlation between two bare modes reaches maximum at Δω=0 (the peak of the green curve in Fig. 2E appears at t=10ms). The corresponding coefficient of each component for $N_+$ is illuminated in Fig. 2F.

In an ideal Otto cycle, the sweep time should be fast enough to avoid the dissipation of membranes, and at the same time slow enough to avoid the diabatic transition between two normal modes, in order to keep the phonon number of the upper normal mode constant. In the real situation, the dissipation cannot be completely ignored during this stroke. Following the previous theoretical proposals, we still adopt the term of "adiabatic", and exploit a more realistic model, Landau-Zener model [50-53], to characterize the transitions including the dissipation in such a process. The Landau-Zener transition depends on the frequency sweep rate α and the coupling strength Λ. When the frequency sweep time is slow enough, the phonon population transfer from $M_1$ to $M_2$ adiabatically, i.e. along the upper normal mode branch. When the sweep time is fast, some population can transfer to the lower normal branch, i.e. diabatic. The diabatic transition probability can be approximately expressed as $P_{diab} = \exp[-2\pi\Lambda^2/|\alpha|]$ [50]. In Fig. 2, Λ=2π×40 Hz and |α|=2π×20 Hz/ms, which leads to $P_{diab}$≈4%. Therefore, the transfer is approximately an adiabatic transition, while the phonon number of the upper normal mode dissipates during the transition. The oscillation in Fig. 2C is the Rabi oscillation in the non-resonant case, i.e. with a frequency detuning.

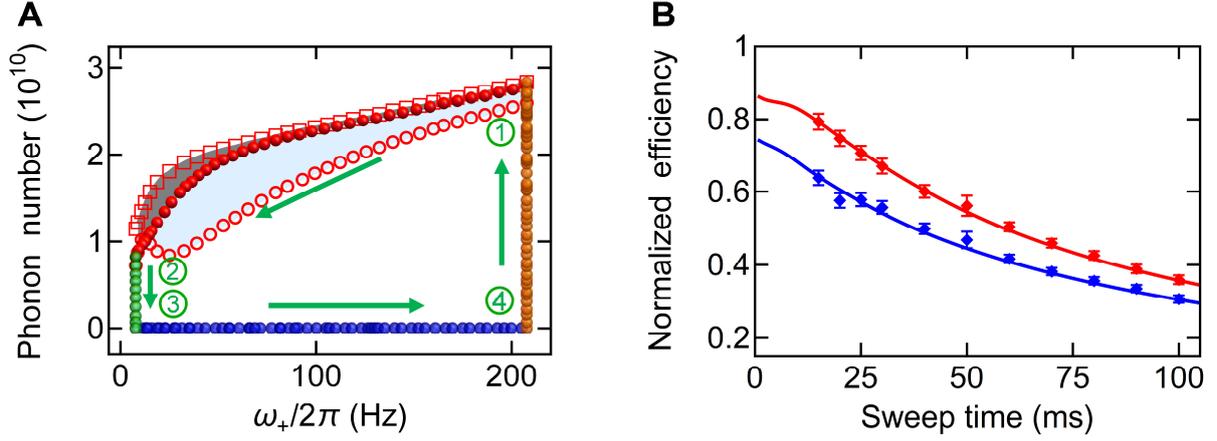

**Fig. 3. Thermodynamic cycle and work efficiency.** (A) The thermodynamic cycle of engine with the upper normal mode. The status numbers are marked according to Fig. 1D. The red dots represent the phonon number of the upper normal mode, the red squares present the total phonon numbers of the upper and lower normal modes, and the red circles are the phonon number of upper normal mode by ignoring the correlations (the cross terms in N₊). The fluctuations of phonon numbers follow the Boltzmann distribution, which are not shown in the figure. (B) The normalized efficiency as a function of the frequency sweep time. The red and blue diamonds are the experimental data with and without the contribution of correlations. The red and blue curves are the corresponding theoretical simulations. The error bars are the standard deviations.

The thermodynamic diagram of the Otto cycle is shown in Fig. 3A. The phonon number of the upper normal mode for four strokes is plotted as a function of the upper normal mode frequency ω₊, with a frequency sweep time of 15 ms, which is more close to an ideal Otto cycle compared to a longer sweep time. The cycle starts from the expansion stroke (the top right corner in Fig. 3A), which corresponds to t=0 in Fig. 2B. The phonon number decays are determined by the rates of $\gamma_1$ and $\gamma_2$ (the red dots in Fig. 3A). Then the cycle follows the processes of thermalization of the normal mode to the low phonon population (the green dots), "adiabatic" compression (the blue dots), and thermalization to the high phonon population state (the orange dots) in the counterclockwise direction as illuminated in Fig. 3A.

Since the phonon number is not conserved in the expansion stroke in Fig. 3A, in contrast to an ideal Otto cycle, the work performed in the expansion stroke should be an integral $W_{1\to 2} = \hbar \int_{\omega_{+,i}}^{\omega_{+,f}} N_+(\omega_+) d\omega_+$. Consequently, the total work W done by the upper normal mode is estimated straightforwardly with the area enclosed by the cycle (the product of two axis units is the energy divided by h), which gives W = 3×10$^{-21}$ J according to Fig. 3A and a power ~ 3.75×10$^{-21}$J/s. Due to a relatively fast sweep rate compared to the situation shown in Fig. 2, i.e. |α|=2π×27 Hz/ms, the diabatic transition probability becomes 9%. The diabatic transition can be visualized by the grey area in Fig. 3A, where the red squares present the total phonon numbers of the upper and lower normal modes. The difference between the red squares and the red dots are due to the diabatic transition from the upper normal mode to the lower normal mode. The red circles are the phonon number of upper normal mode without including the correlations (ignoring the cross terms in $N_+$). Therefore, the light blue area in Fig. 3A can be understood as the work done by the correlations of the bare modes.

In general, the injected heat has an integral expression similar to the work calculation. In the current setup, the frequency doesn't change during the thermalization process, which is shown by the orange dots in Fig. 3A, therefore, $Q_{4\rightarrow 1} = \hbar\omega_{+,i}(\langle N_+\rangle_i - \langle N_+\rangle_f) \approx \hbar\omega_{+,i}\langle N_+\rangle_i$. Since the coupling strength Λ=2π×40 Hz is much smaller than the mechanical mode frequency $\omega_m \approx$ 2π×400 kHz, the local approach can be adopted [37]. The work efficiency is limited by the difference of the initial and final normal mode frequency in stroke 1 [29]. The work efficiency $\eta_{ideal}$ is approximately equal to $(\omega_{+,i}-\omega_{+,f})/\omega_m \approx$ 0.05% for an ideal Otto cycle (with negligible small decays and the Otto cycle has a rectangular shape) in the current setup, where $\omega_{+,i}$ and $\omega_{+,f}$ (y axis in Fig. 1D) are the initial and final frequencies of the upper normal mode with a frequency detuning $\Delta\omega_{i,f}$ (x axis in Fig. 1D). A larger value of $\omega_{+,i}-\omega_{+,f}$ could be chosen to increase the work efficiency. A significant enhancement of the efficiency requires a much stronger deformation of the membrane, for example, by using a much larger voltage applied on the piezo or using a more efficient piezo actuator. There is no limiting factor, since the frequency difference $\omega_i-\omega_f$ can be as large as $\omega_m$ in principle.

Here, the normalized efficiency $\eta_N$ is used to focus on the effects of diabatic transition and decays, which is defined as the ratio of the actual work efficiency to $\eta_{ideal}$, i.e., $\eta_N=W/(\eta_{ideal} Q_{4\rightarrow 1})$. $\eta_N$ as a function of the frequency sweep time is presented in Fig. 3B. $\eta_N$ decreases with the increase of the sweep time, since the phonon number damping is dominant (the red diamonds and red curve). $\eta_N$ without including the correlations (the cross terms in $N_+$) is shown as the blue diamonds and blue curve. When the sweep time is relatively small, the diabatic transfer occurs. This issue could be solved by using shortcuts to adiabaticity, which allows the fast control of heat engine without decreasing the efficiency [54,55].

The contribution of the lower normal mode is much smaller than the upper normal mode. This is because the upper and lower normal modes are respectively dominated by the bare modes $M_1$ and $M_2$ at $\Delta\omega_i \sim 2\pi\times200$ Hz, and consequently the upper normal mode is thermalized to a high phonon population and the lower normal mode is thermalized to a low phonon population, with a ratio of $\sim 60$. Therefore, the work efficiencies for the cases of the upper normal mode and two normal modes are similar. While the situation is different at a relatively small sweep time, the efficiency of two normal modes could have a significantly greater value owing to the diabatic transition. The thermodynamic cycle of the lower normal mode and more details can be found in the supplementary materials.

Previously, we have demonstrated a single cylinder optomechanical heat engine. In practice, a multi-cylinder heat engine is preferred, due to the fact that not only more work can be performed in each cycle, but also the work can be extracted more smoothly. Here we realize a straight-twin nanomechanical heat engine in the same setup but engineering the normal modes in a different way, which allows two normal modes to operate out-of-phase as two cylinders. The schematic of Otto cycle for the straight-twin engine is shown in Fig. 4A, with both normal modes included (the notation in Fig. 4A is defined in the same way as in Fig. 1B). The normal mode branches is modified to be $\omega_\pm = \pm\sqrt{\Delta\omega^2/4 + \Lambda^2}$, as shown in Fig. 4B. In contrast to the situation in Fig. 1D, where $\omega_1$ changes and $\omega_2$ keeps constant, here both frequencies of membranes changes at the same sweep rate but with opposite directions, and then one can observe a noise spectrum as shown in Fig. 4B. By following the control sequence shown in Fig. 4C, the phonon numbers of $M_1$ and $M_2$ during a cycle are measured (Fig. 4D). In order to realize positive work output for both normal modes, it requires that the sweep frequency separation range to be offset towards the right from the anti-crossing [36], e.g. $\Delta\omega$ sweeps from $2\pi\times720$ Hz to $-2\pi\times180$ Hz, and then sweeps back, as shown in Fig. 4B. Similar to the case in a single cylinder heat engine, one can obtain the thermodynamic diagrams of both normal modes in one cycle, as shown in Figs. 4E and 4F. The top right corner in Fig. 4E and the left bottom corner in Fig. 4F correspond to the initial time of a cycle, i.e. t=0. The red, green, blue, and orange dots represent the four strokes in sequence. The work done by the two normal modes in one thermodynamic cycle are $2.6\times10^{-21}$ J and $0.3\times10^{-21}$ J, respectively. The difference of the work performed by two normal modes can be reduced by using the eigenmodes of membranes with larger quality factors.

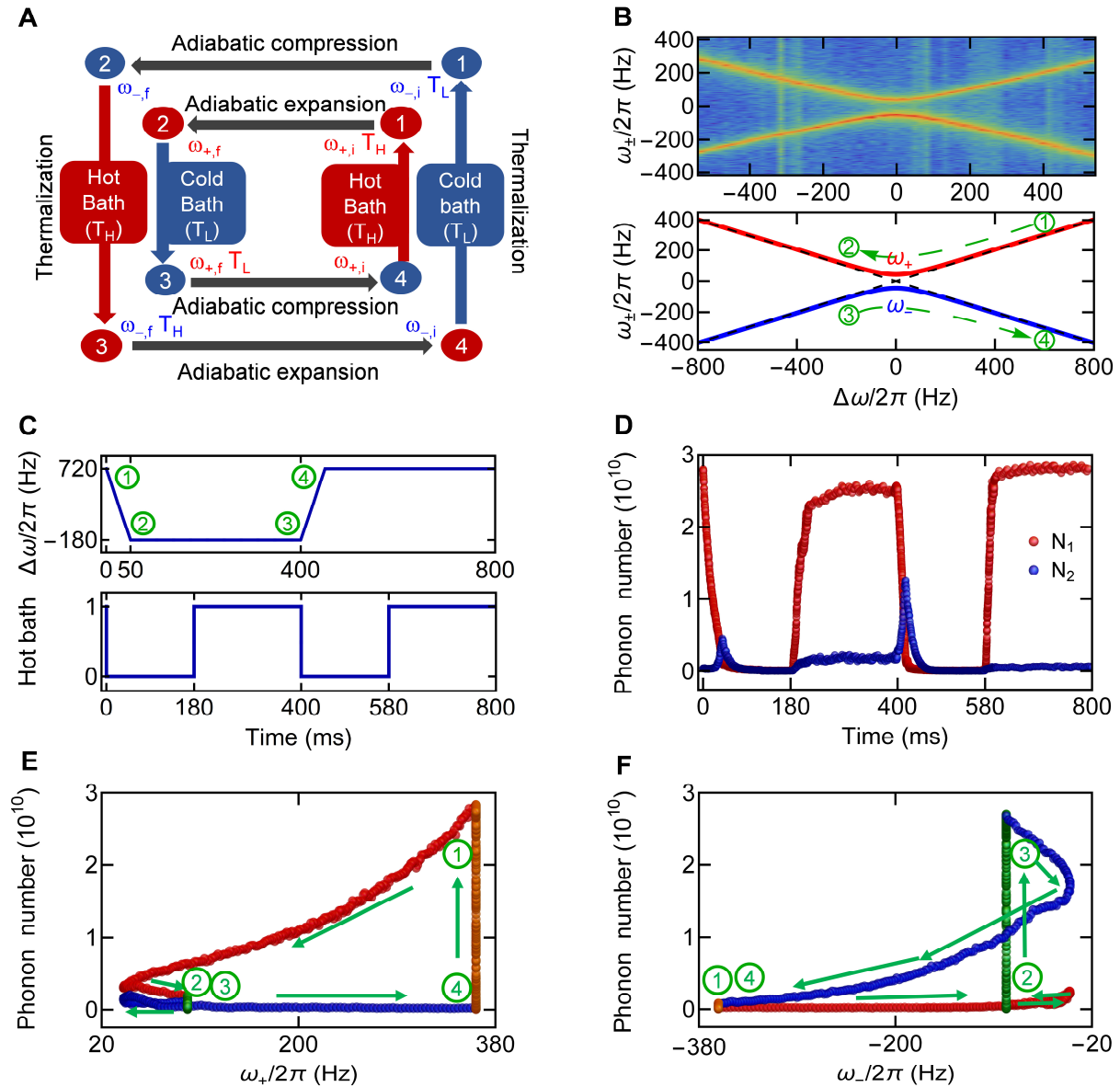

**Fig. 4. Straight-twin nanomechanical heat engine.** (A) Schematic of the Otto cycle for a straight-twin engine with both normal modes. (B) The experimental measurement and the theoretical simulation of the normal mode frequencies for a straight-twin engine. (C) The frequency detuning and the high temperature thermal bath connected to $M_1$ as a function of time during a cycle. (D) The measured average phonon numbers of $M_1$ and $M_2$ during a cycle. (E,F) The thermodynamic diagrams of the upper and lower normal mode branches, respectively. The status numbers are marked according to Fig. 4B. The fluctuations of phonon numbers follow the Boltzmann distribution, which are not shown in the figures.

In conclusion, we have realized a coupled-mode stochastic heat engine with nanomechanical resonators. The heat engine is based on a multimode optomechanical system, and the normal mode of two coupled nanomechanical membranes is used as the working medium. Both single and double-cylinder heat engines are experimentally investigated by properly engineering the normal modes of the coupled system. It is a truly multiple cylinder heat engine scheme and possible to generalize to a heat engine array with sophisticated engineering and fabrication. Future work includes investigating the finite-time quantum thermodynamics, and enhancement of the output power and efficiency in such a novel platform for the heat engines at nanoscales.

## MATERIALS AND METHODS

**Nanomechanical membrane**

The nanomechanical membrane used in the experiment is a commercial stoichiometric silicon nitride membrane (Norcada), which is deposited on a silicon wafer by low-pressure chemical vapor deposition (LPCVD). It has a large tensile stress ~ 1 GPa, a high mechanical quality factor (~ $10^7$ at low temperature), and ultralow optical absorption in the infrared range. The membrane has a thickness of 50 nm and a 1×1 mm$^2$ size. Each membrane is attached to two ring piezo actuators (Noliac) for the purpose of independent membrane frequency and position control.

**Acknowledgments:** We thank Keye Zhang for helpful discussions. **Funding:** This research was supported by the National Key R&D Program of China (No. 2017YFA0304201), NSFC (No. 11925401, No. 11734008, No. 11974115, No. 11621404), the Shanghai Municipal Science and Technology Major Project (No. 2019SHZDZX01), the Program for Professor of Special Appointment (Eastern Scholar) at Shanghai Institutions of Higher Learning. **Author contributions:** J.S., C.Y., and H.W. carried out the experiment, analyzed the data, developed the theory, and wrote the paper. **Competing interests:** The authors declare no competing interests. **Data and materials availability:** All data needed to evaluate the conclusions in the paper are present in the paper and/or the Supplementary Materials.